\documentclass[aps,pra,twocolumn,showpacs,superscriptaddress,10pt]{revtex4-1}
\usepackage{amsmath,amsfonts,amssymb}
\usepackage{xcolor,graphicx,hyperref,epstopdf}
\usepackage{braket,ulem}

\newtheorem{corollary}{Corollary}

\begin{document}
\title{Quantifying nonclassicality by characteristic functions}

\author{S. Ryl}\email{sergej.ryl@uni-rostock.de}
\affiliation{Arbeitsgruppe Theoretische Quantenoptik, Institut f\"ur Physik, Universit\"at Rostock, D-18051 Rostock, Germany}
\author{J. Sperling}
\affiliation{Clarendon Laboratory, University of Oxford, Parks Road, Oxford OX1 3PU, United Kingdom}
\author{W. Vogel}
\affiliation{Arbeitsgruppe Theoretische Quantenoptik, Institut f\"ur Physik, Universit\"at Rostock, D-18051 Rostock, Germany}

\date{\today}

\begin{abstract}
	In this paper, we use the characteristic function, i.e., the Fourier transform of the Glauber-Sudarshan phase-space distribution, to determine the degree of nonclassicality of a given state.
	This degree of nonclassicality quantifies the nonclassicality in terms of quantum superpositions.
	We demonstrate two ways to exactly find or to lower-bound the degree of nonclassicality by studying the properties of the characteristic functions.
	The developed criteria are applied to two examples of squeezed states undergoing a classical mixing or a nonclassical superposition with vacuum.
\end{abstract}


\maketitle

\section{Introduction.}
	The quantum superposition principle leads to observations that are incompatible with classical physics.
	These quantum phenomena include, for example, sub-Poisson photon-number statistics~\cite{Mandel1979} or quadrature variances below the vacuum level~\cite{Slusher1985}.
	The general nonclassicality of a single-mode light field is defined in terms of the Glauber-Sudarshan $P$ function~\cite{Glauber1963,Sudarshan1963,Titulaer1965,Mandel1986}.
	However, this definition only separates classical (coherent states and classical statistical mixtures of them) and nonclassical states, without quantifying the amount of nonclassicality.

	There are several ways to quantify the nonclassicality.
	One possible measure is the distance (based on the trace norm) between the state under study and the closest classical state, which was introduced by Hillery~\cite{Hillery1987a}.
	It was shown to be, in general, not computable due to an infinite number of parameters to be optimized~\cite{Asboth2005}.
	Recently, this measure was generalized for multimode states and calculated for some finite-dimensional examples~\cite{Nair2017}.
	Other distance-based nonclassicality measures are described in Refs.~\cite{Wuensche2001,Dodonov2000,Marian2002,Marian2004}.
	However, such approaches to quantify the nonclassicality are ambiguous as they strongly depend on the chosen norms or metrics defining the distances~\cite{Sperling2015}.

	Another quantifier of nonclassicality is based on the convolution of the $P$~function with the amount of thermal noise needed to get a non-negative phase-space function~\cite{Lee1991}.
	In fact, Lee's so-called nonclassical depth basically determines the robustness of a nonclassical state against thermal noise.
	The connections between such a robustness and negativity-based entanglement monotones have also been studied~\cite{Miranowicz2015}.
	It is noteworthy that the nonclassical depth can be discontinuous when the quantum state undergoes tiny changes~\cite{Luetkenhaus1995}.
	Another nonclassicality measure is based on negativities of the Wigner function~\cite{Kenfack2004}.
	However, negativities of the Wigner function do not appear for all nonclassical states.
	For example, squeezed states are nonclassical, but they have a positive Gaussian Wigner function.

	Other measures are based on the entanglement potential of nonclassical states~\cite{Asboth2005}, making use of the connection between single-mode nonclassicality and bipartite entanglement~\cite{Aharonov1966,Kim2002,Wang2002}.
	However, such an approach maps the problem of quantifying nonclassicality to the at least equally cumbersome problem of quantifying entanglement.
	For example, from the two suggested measures in Ref.~\cite{Asboth2005}, the first one solely addresses the partial transpositions and the second one requires an optimization over all two-mode separable states.
	Recently, a specific nonclassicality measure for general two-mode Gaussian states has also been developed~\cite{Miranowicz2016,Arkhipov2016}.

	Besides the previously mentioned nonclassicality quantifiers based on distances, robustness or entanglement, a different, algebraic way to quantify nonclassicality has been developed~\cite{Gehrke2012}.
	This so-called degree of nonclassicality is based on the fundamental quantum superposition principle.
	It quantifies the amount of nonclassicality in terms of the minimal number of coherent states that are needed to be superimposed in order to represent the state under study and it is a member of a general class of algebraic measures, applying to different notions of nonclassicality~\cite{Sperling2015}.
	Also, the degree of nonclassicality has the advantage of being directly related~\cite{Vogel2014} to the quantification of entanglement through the Schmidt number~\cite{Sanpera2001,Ternal2000}.
	Furthermore, a moment-based approach was introduced to formulate measurable witnesses for the degree of nonclassicality~\cite{Mraz2014}.
	Such a witness approach allows for the experimental verification of the amount of nonclassicality.
	More precisely, it ensures the least degree of nonclassicality which can be certified under given experimental conditions.

	In the present article, we formulate an approach to determine the degree of nonclassicality based on the Fourier transform of the Glauber-Sudarshan $P$~function, the characteristic function.
	As the characteristic function contains all information about the state, including its degree of nonclassicality, we are elaborating two ways to construct conditions for determining the degree of nonclassicality.
	The first way is based on the analysis of polynomial characteristic functions which allows us to infer the degree of nonclassicality for a specific class of states.
	The second way is related to the application of Bochner's theorem~\cite{Bochner1933} in quantum optics~\cite{Vogel2000,Richter2002}, which is not limited to a specific class of states.

	We organized this work as follows.
	In Sec.~\ref{sec:doN}, we motivate our choice of the degree of nonclassicality, we give the definition, and we list some examples.
	In Sec.~\ref{sec:finSpaces}, we derive our first results for the degree of nonclassicality for polynomial characteristic functions.
	Section~\ref{sec:Boch} provides our second method to estimate the degree of nonclassicality for arbitrary states.
	We summarize the results in Sec.~\ref{sec:Conc}.

\section{The degree of nonclassicality}\label{sec:doN}

\subsection{Motivation}
	Let us motivate why we use the degree of nonclassicality (DNC) as our figure of merit.
	The classical electric field strength can be recovered as the expectation value of the field operator in a coherent state.
	The coherent states carry the same minimal quantum noise as the vacuum state.
	This minimal noise level is necessary to fulfill the Heisenberg uncertainty relation, resulting from the noncommutativity of the creation and annihilation operators.
	In this sense, the quantum states most similar to a classical coherent field are the coherent states, which are often simply denoted as (pure) classical states.
	As long as a state under study can be represented as a classical mixture of coherent states, it is also called classical~\cite{Titulaer1965,Mandel1986}.
	If a given quantum state cannot be represented as a classical mixture of coherent states, it must contain quantum superpositions of coherent states.

	The property of a state either being classical or not can have only two discrete truth values, ``yes'' or ``no''.
	Consequently, even infinitesimal changes of the density matrix of a given state may switch its property between classical and nonclassical.
	Of course, in an experiment, only significant (with respect to the error bars) changes of the state can certify transitions between classical and nonclassical properties.

	The fundamental concept underlying the nonclassicality of a quantum states is the quantum superposition principle.
	This leads to quantum interference effects which have no counterpart in classical physics.
	As quantum superpositions are the origin of nonclassicality, it is natural to quantify them in terms of the number of quantum superpositions needed to represent a given state.
	Similarly to the binary property of being classical or nonclassical, the property that a state includes a certain number of quantum superpositions also attains discrete values.
	This relates to a larger alphabet of truth values which can be used for encoding quantum information.

	The DNC was introduced in Ref.~\cite{Gehrke2012}.
	More generally, it was shown that the measure of nonclassicality in terms of superpositions requires only two properties~\cite{Sperling2015}:
	First, the DNC attains the smallest value for classical states.
	Second, the DNC does not increase under general classical operations.
	As the single-mode nonclassicality is connected to bipartite entanglement~\cite{Aharonov1966,Kim2002,Wang2002}, such a link should exist between the corresponding measures.
	In fact, the DNC exactly agrees with the Schmidt-number measure of the entanglement in the output channels when a nonclassical single-mode state is superimposed with the vacuum state on a beam splitter~\cite{Vogel2014}.

	Therefore, the DNC presents a nonclassicality measure that is based on the fundamental axiom of quantum superposition.
	It generalizes the binary answers ``yes'' or ``no'' to the question of classicality of a state to discrete levels of the amount of nonclassicality.
	Moreover, it has a unique one-to-one relation to the Schmidt-number measure of entanglement.

\subsection{Definition and Examples}
	The Glauber-Sudarshan representation~\cite{Glauber1963,Sudarshan1963} of a single-mode state $\hat{\rho}$ reads
	\begin{equation}
		\hat{\rho}=\int d^{2}\alpha P(\alpha)|\alpha\rangle\langle\alpha|, \label{eq:GS}
	\end{equation}
	whereby one also defines the so-called $P$~function.
	If $P$ is a classical probability density, the state is called classical.
	If there is no such classical representation of the state, then the state is a nonclassical one.
	The nonclassicality of a state stems from the fact that quantum interferences are needed to describe such a nonclassical state~\cite{Sperling2015}.
	One can separate quantum effects and classical statistics by expanding a state in terms of superpositions of coherent states,
	\begin{equation}
		|\Psi_{r}\rangle=\lambda_{1}|\alpha_{1}\rangle+\dots+\lambda_{r}|\alpha_{r}\rangle, \label{eq:Psir}
	\end{equation}
	with $\alpha_{i}\neq\alpha_{j}$ for $i\neq j$, and complex values $\lambda_{i}\neq0$ for $i=1,\dots, r$.
	The smallest number $r$ providing the representation of a pure state in the form~\eqref{eq:Psir} defines the DNC of this state~\cite{Gehrke2012},
	\begin{equation}
		D_{\rm ncl}(|\Psi_{r}\rangle)=r.
	\end{equation}
	For example, any classical coherent state consists of a single element in the expansion~\eqref{eq:Psir} and, therefore, has the minimal DNC of~1, $D_{\rm ncl}(|\alpha\rangle)=1$.

	For mixed states, we define the set $\mathcal{S}_{r}$ being the closure of the convex combination of all pure states with a number of superimposed coherent states being less than or equal to $r$, $\mathcal S_r=\overline{\mathrm{conv}\{|\Psi_{r'}\rangle\langle\Psi_{r'}|:D_{\rm ncl}(|\Psi_{r'}\rangle)\leqslant r\}}$.
	With other words, the DNC of a mixed state $\hat{\rho}$ is the minimal number $r$ for which this state can be written as a classical mixture
	\begin{equation}
		\hat{\rho}=\int d P_{\rm cl}(|\Psi_{r}\rangle)|\Psi_{r}\rangle\langle\Psi_{r}|, \label{eq:rmix}
	\end{equation}
	with a classical probability distribution $P_{\rm cl}$ over $\mathcal S_r$~\cite{Mraz2014}.
	It also directly follows for states with DNC greater than~$r$ that $P_{\rm cl}$ can be replaced with a quasiprobability distribution that has negativities.
	In fact, for $r=1$, the latter reduces to the Glauber-Sudarshan $P$ function.
	Our quantifier of the nonclassicality of a state $\hat\rho$ is the minimal number $r$ for which the decomposition in Eq.~\eqref{eq:rmix} is possible,
	\begin{equation}
		D_{\rm ncl}(\hat\rho)=r,
	\end{equation}
	being equivalent to $\hat\rho\in\mathcal S_{r}\setminus\mathcal S_{r-1}$.

	Let us list some states with known DNC~\cite{Vogel2014}.
	The DNC of Fock or photon number states $|n\rangle$ reads
	\begin{equation}
		D_{\rm ncl}(|n\rangle)= n+1, \label{eq:DNC_Fock}
	\end{equation}
	and for the displaced squeezed states $|\xi;\alpha\rangle$ holds
	\begin{equation}
		D_{\rm ncl}(|\xi;\alpha\rangle)=\infty. \label{eq:DNC_sq}
	\end{equation}
	These results are helpful and are discussed in the next sections.
	It is also worth pointing out that any state can be expanded in terms of coherent states and has therefore a well-defined DNC.

\section{Characteristic Functions in Finite Spaces}\label{sec:finSpaces}
	In this section, we derive a method to determine the DNC for states with a finite Fock-state expansion.
	That is, the characteristic function (CF) is a polynomial.
	First, we find the DNC of a finite superposition of Fock states and for mixtures of such states.
	Then, we connect polynomial CFs with the finite Fock-state expansion.
	This connection yields a method to determine the DNC from the polynomial CF.
	Afterwards, we study the influence of photon addition on the CFs of states with finite Fock-state expansion.
	Eventually, we discuss our results in relation to other nonclassicality quantifiers.

\subsection{Finite superpositions of Fock states}
	It is well known that states with a finite Fock-state expansion---excluding the vacuum state---are nonclassical states.
	In terms of quantifying nonclassicality, Fock states themselves obey Eq.~\eqref{eq:DNC_Fock}.
	The proof of Eq.~\eqref{eq:DNC_Fock} may be separated into two parts.
	First, $D_{\rm ncl}(|n\rangle)\leqslant n+1$ can be shown by representing the state $|n\rangle$ as a superposition of $n+1$ coherent states.
	Second, $D_{\rm ncl}(|n\rangle)\geqslant n+1$ is required since the state after beam-splitter transformation has Schmidt number~$n+1$.

	In order to derive the DNC of a finite superposition of Fock states, we need two equations from Ref.~\cite{Vogel2014}.
	The first one allows one to represent the Fock state $|n\rangle$ as a superposition of coherent states,
	\begin{equation}
		|n\rangle=\lim\limits_{\alpha\to 0}\lim\limits_{\varepsilon\to 0}\sum_{j=0}^n \frac{\sqrt{n!}(-1)^{n-j}}{j!(n-j)!}\frac{e^{|\alpha+j\varepsilon|^2/2}}{\varepsilon^{n}}|\alpha+j\varepsilon\rangle,\label{eq:Fock_via_coh}
	\end{equation}
	which is a decomposition in terms of $n+1$ coherent states $|\alpha+j\varepsilon\rangle$.
	The second equation shows the connection between a single-mode Fock state $|n\rangle$ and a two-mode state with Schmidt rank $n+1$.
	Combining Fock state $|n\rangle$ and the classical vacuum state $|0\rangle$ on a 50:50 beam splitter (BS) results in the two-mode state
	\begin{equation}
		|n\rangle\otimes|0\rangle=|n,0\rangle\,\stackrel{\rm BS}{\longmapsto}\, \frac{1}{2^{n/2}}\sum_{j=0}^n \binom{n}{j}^{1/2} |j,n-j\rangle,
	\end{equation}
	which has a Schmidt rank of $n+1$.
	Now, let us consider the finite superposition of up to $m$ photons,
	\begin{equation}\label{eq:sup_Fock}
		|\psi\rangle=\sum\limits_{n=0}^{m}\lambda_n|n\rangle, \text{~with }\lambda_m\neq0.
	\end{equation}
	Inserting Eq.~\eqref{eq:Fock_via_coh} and rearranging the sums---which can be done as all sums are finite---we get
	\begin{equation}\label{eq:genFockBS}
		|\psi\rangle=\lim\limits_{\alpha\to 0}\lim\limits_{\varepsilon\to 0}\sum\limits_{j=0}^{m}\sum\limits_{n=j}^{m}\lambda_{n}\frac{\sqrt{n!}(-1)^{n-j}}{j!(n-j)!\varepsilon^n}e^{|\alpha+j\varepsilon|^2/2}|\alpha+j\varepsilon\rangle.
	\end{equation}
	Thus, maximally $m+1$ coherent states are needed to represent the state $|\psi\rangle$, so $D_{\rm ncl}(|\psi\rangle)\leqslant m+1$.

	We can split this state by a $50{:}50$ beam splitter.
	This results in
	\begin{eqnarray}
		|\psi,0\rangle
		&\stackrel{{\rm BS}}{\longmapsto}&
		\sum\limits_{n=0}^{m}\frac{\lambda_n}{2^{n/2}}\sum\limits_{j=0}^{n}\binom{n}{j}^{1/2}|j,n-j\rangle
		\\\nonumber &&=
		\sum\limits_{j=0}^{m}|j\rangle\otimes\left[\sum\limits_{n=j}^{m}\frac{\lambda_n}{2^{n/2}}\binom{n}{j}^{1/2}|n-j\rangle\right].
	\end{eqnarray}
	Since $\lambda_m\neq0$ there are $m+1$ linearly independent states needed to represent the two-mode state, which implies a Schmidt number of $m+1$.
	Hence, the DNC of the one-mode state $|\psi\rangle$ must fulfill $D_{\rm ncl}(|\psi\rangle)\geqslant m+1$.
	In combination with the inequality from the previous paragraph, $D_{\rm ncl}(|\psi\rangle)\leqslant m+1$, we find
	\begin{equation}
		D_{\rm ncl}(|\psi\rangle)=m+1.
	\end{equation}
	Thus, we have shown that for any finite superposition of Fock states in Eq. \eqref{eq:sup_Fock} holds that the highest photon-number contribution defines the DNC of such states.
	
	We can also generalize this method for finite spaces to mixed states.
	Suppose we have a density operator of the form
	\begin{align}\label{eq:mixFock}
		\hat\rho=\sum_{k,l=0}^m \rho_{k,l} |k\rangle\langle l|, \text{ with }\rho_{m,m}\neq0.
	\end{align}
	As this state has a spectral decomposition in terms of different states $|\psi_s\rangle$ of the form \eqref{eq:sup_Fock} together with
	\begin{equation}
		\hat{\rho}=\sum_{s} p_s|\psi_s\rangle\langle\psi_s|,
		\text{ with } p_s\geqslant 0
		\text{ and } \sum_{s}p_s=1, \label{eq:decomp_fin_fock}
	\end{equation}
	we can conclude from convexity that \mbox{$D_\mathrm{ncl}(\hat\rho)\leqslant m+1$.}
	Moreover, any possible convex decomposition~\eqref{eq:decomp_fin_fock} requires that at least one state $|\psi_s\rangle$ has a nonvanishing contribution of $|m\rangle$.
	Hence, this $|\psi_s\rangle$ has a DNC of $m+1$ as we have shown above.
	From the construction of the DNC in terms of convex hulls of pure states [Eq.~\eqref{eq:rmix}], we get
	\begin{align}
		D_\mathrm{ncl}(\hat\rho)=m+1.
	\end{align}
	Thus, any pure or mixed state with a finite photon number has a DNC which is given by $m+1$, where the highest contributing photon number is $m$.

\subsection{Polynomial CFs}\label{subsec:PolyCF}
	The CF $\Phi(\beta)$ is the Fourier transform of the $P(\alpha)$~function and it can be written as the normal-ordered expectation value of the displacement operator $\hat D(\beta)=\exp(\beta\hat a^\dag-\beta^\ast\hat a)$,
	\begin{align}
		\Phi(\beta)=\langle{:}\hat D(\beta){:}\rangle=\int d^2\alpha P(\alpha)e^{\beta\alpha^\ast-\beta^\ast\alpha}.
	\end{align}
	Say the characteristic function is a $2m$-th order polynomial of $\beta$ in the form
	\begin{align}\label{eq:polynomialCF}
		\Phi(\beta)=\sum_{k,l=0}^m \phi_{k,l} \beta^{k}\beta^{\ast l}.
	\end{align}
	Note that $\phi_{k,l}=(-1)^l\langle\hat{a}^{\dag k}\hat{a}^{l}\rangle/(k!l!)$.
	The maximal sum of powers of $\beta$ and $\beta^\ast$ is referred to as the order of the polynomial.
	We may compare this with a Fock matrix element $|k\rangle\langle l|$ from Eq.~\eqref{eq:mixFock}, having the CF~\cite{Sperling2016}
	\begin{align}
		\Phi_{k,l}(\beta)=\langle l|{:}\hat D(\beta){:}|k\rangle
		=\sum_{n=0}^{\min\{k,l\}}\frac{\sqrt{k!l!}\beta^{l-n}(-\beta^\ast)^{k-n}}{n!(k-n)!(l-n)!},
	\end{align}
	which is a polynomial of the $(k+l)$th order.
	For instance, the Fock state $|m\rangle$ yields a polynomial CF $\Phi_{m,m}(\beta)=L_m(|\beta|^2)$, with the $m$th Laguerre polynomial~$L_m$.
	In general, the CF is a polynomial of order $2m$ for states with a finite Fock expansion~\eqref{eq:mixFock}.
	Note that in terms of the $P$ function, this means that $P$ has singularities of the type of a $2m$th derivative of the $\delta$ distribution~\cite{Cahill1969}.
	Conversely, any $P$ function with a finite order $2m$ of derivatives of $\delta$ distributions has a CF of the form of Eq.~\eqref{eq:polynomialCF}.

	In summary, nonclassical quantum states with a finite Fock expansion \eqref{eq:mixFock} have a DNC of $m+1$ when the maximally required Fock state is $|m\rangle$.
	If and only if the state is of this form, the CF is a $2m$-th order polynomial.
	This result implies the following corollary:
	\begin{corollary}
		If the CF is a polynomial of order $2m$, then the state has DNC of $m+1$.
	\end{corollary}
	It is worth mentioning that $m=0$ describes a vacuum state---a classical coherent state with a coherent amplitude of zero.
	Vacuum is described by a minimal DNC of~1 and a CF that is a constant function.
	In the continuation of this section, we describe the photon addition as a nonclassical process~\cite{Rahimi2013}.
	Its impact on the polynomial CFs is analyzed.

\subsection{Photon addition}\label{sec:PhAdd}
	Fock states can be regarded as photon-added vacuum states, $\sqrt{n!}|n\rangle=\hat a^{\dag n}|0\rangle$.
	In this case, each operation of photon creation (also denoted as photon addition) increases the DNC of the state by~1.
	Considering $m$ iterations of such a process, we get
	\begin{eqnarray}
		&&D_{\rm ncl}(|n+m\rangle)=D_{\rm ncl}(|n\rangle)+m.
	\end{eqnarray}
	In this section, we study the influence of photon addition on arbitrary states in terms of CFs and generalize the above equation.

	For a general density operator, an $n$-photon-addition process reads as
	\begin{equation}\label{eq:multiadd}
		\hat{\rho}_{\rm out}=\mathcal{N}\hat{a}^{\dag n}\hat{\rho}_{\rm in}\hat{a}^n,
	\end{equation}
	where $\mathcal{N}=[{\rm tr}(\hat{a}^{\dag n}\hat{\rho}_{\rm in}\hat{a}^n)]^{-1}$ is the normalization constant.
	Examples of experimental realizations of these processes are described in Refs.~\cite{Bellini2007,Zavatta2007,Kiesel2008} and additional details on the theory can be found in Refs.~\cite{Rahimi2013,Fiurasek2009,Dodonov2009,Sperling2014}.
	For a coherent state, we can write
	\begin{align}\label{eq:addedcoherentstate}
		\hat a^{\dag n}|\alpha\rangle=e^{-|\alpha|^2/2}\partial_\alpha^n e^{|\alpha|^2/2}|\alpha\rangle=e^{-|\alpha|^2/2}\partial_\alpha^n e^{\alpha\hat a^\dag}|0\rangle.
	\end{align}
	Writing both the input and the output density operators in the Glauber-Sudarshan representation,
	$\hat\rho_x=\int d^2\alpha P_x(\alpha)|\alpha\rangle\langle\alpha|$ ($x\in\{\mathrm{in},\mathrm{out}\}$),
	we find after some algebra the input-output relation for the $P$~functions.
	That is, we get from Eq. \eqref{eq:addedcoherentstate}
	\begin{align}\nonumber
		\hat\rho_\mathrm{out}=&\mathcal N\int d^2\alpha P_\mathrm{in}(\alpha)e^{-|\alpha|^2}\partial_\alpha^n\partial_{\alpha^\ast}^n\left[e^{|\alpha|^2}|\alpha\rangle\langle\alpha|\right]
		\\=&\int d^2\alpha \mathcal N e^{|\alpha|^2}\partial_\alpha^n\partial_{\alpha^\ast}^n\left[P_\mathrm{in}(\alpha)e^{-|\alpha|^2}\right]|\alpha\rangle\langle\alpha|,
	\end{align}
	which results in
	\begin{equation}
		P_{\rm out}(\alpha)=\mathcal{N}e^{|\alpha|^2}\partial_{\alpha}^n\partial_{\alpha^{\ast}}^n\left[ P_{\rm in}(\alpha)e^{-|\alpha|^2}\right].
	\end{equation}
	
	By performing the Fourier transform of the $P$ function, we find the influence of photon addition on the CF, $\Phi$:
	\begin{eqnarray}
		\nonumber &&\Phi_{\rm out}(\beta)=\int{d}^2\alpha\,e^{\alpha^{\ast}\beta-\beta^{\ast}\alpha} P_{\rm out}(\alpha)\\
		\nonumber &=&\mathcal{N}\int{d}^2\alpha \left[\partial_{\alpha}^n\partial_{\alpha^\ast}^n{e}^{\alpha^\ast\beta-\beta^{\ast}\alpha+\alpha^{\ast}\alpha}\right]P_{\rm in}(\alpha){\rm e}^{-|\alpha|^2}\\
		\nonumber &=&\mathcal{N}\sum\limits_{k=0}^{n}\frac{n!^2}{k!^2(n-k)!}\\
		\nonumber &&\times\int{d}^2\alpha(\beta+\alpha)^k(-\beta^\ast+\alpha^\ast)^k{e}^{\alpha^\ast\beta-\beta^\ast\alpha}P_{\rm in}(\alpha)\\
		\nonumber &=&\mathcal{N}\sum\limits_{k=0}^{n}\frac{n!^2}{k!^2(n-k)!}\\
		\label{eq:Phi_add} &&\times(-|\beta|^2+\beta\partial_{\beta}+\beta^\ast\partial_{\beta^\ast}-\partial_{\beta}\partial_{\beta^\ast})^k\Phi_{\rm in}(\beta).
	\end{eqnarray}
	
	Note that the zero-order derivative is defined as the function itself.
	Equation~\eqref{eq:multiadd} indicates that an input state with a finite Fock expansion up to $m$ photons results in an expansion in photon-number states up to $m+n$.
	Analyzing Eq.~\eqref{eq:Phi_add}, we see that if the input CF is a $2m$-th order polynomial, then the addition of $n$ photons increases the order of the polynomial by $2n$.
	In conclusion, we get the following:
	If the CF $\Phi(\beta)$ of the state $\hat{\rho}_\mathrm{in}$ is a polynomial of order $2m$, then the DNC of an $n$-photon-added state is
	\begin{equation}
		D_{\rm ncl}(\hat{\rho}_\mathrm{out})= m+n+1. \label{eq:res1}
	\end{equation}
	This rule is by design applicable to Fock states and their finite superpositions, which includes all states in finite-dimensional Hilbert spaces. 
	Infinite-dimensional states $\hat\rho$ with $D_{\rm ncl}(\hat\rho)=\infty$, however, remain infinite dimensional under photon addition.

	For comparison, let us also study the photon subtraction.
	The photon annihilation (subtraction),
	\begin{equation}
	  \hat\rho_{\rm out}=\frac{\hat{a}\hat{\rho}_{\rm in}\hat{a}^\dag}{{\rm tr}(\hat{\rho}_{\rm in}\hat{a}^\dag\hat{a})},
	\end{equation}
	is known to be a classical operation~\cite{Gehrke2012}.
	Because of that, it cannot increase the DNC.
	The state of type~\eqref{eq:Psir} is altered by $n$ subtractions to
	\begin{equation}
		|\Psi_r\rangle\mapsto
		\mathcal N\hat{a}^n|\Psi_r\rangle=\mathcal{N}\left(\lambda_1\alpha_1^n|\alpha_1\rangle+\dots+\lambda_r\alpha_r^n|\alpha_r\rangle\right),
	\end{equation}
	where $1/\mathcal{N}^2=\langle\Psi_r|\hat{a}^{\dag n}\hat{a}^n|\Psi_r\rangle$.
	If one of the amplitudes $\alpha_i$ is zero, then the DNC of the state is reduced by~1, otherwise, it remains the same.
	Especially in Eqs.~\eqref{eq:Fock_via_coh} and~\eqref{eq:sup_Fock}, one of the summands vanishes and the DNC for states with finite Fock expansion reduces by $n$ under $n$ photon subtractions.
	For example, the single-photon state $|1\rangle$ with $D_{\rm ncl}(|1\rangle)=2$ is mapped to vacuum state $|0\rangle$ with $D_{\rm ncl}(|0\rangle)=1$ as $\hat{a}|1\rangle=|0\rangle$.
	By contrast to the $n$-photon subtraction, the addition adds $n$ to the DNC of states with a finite Fock expansion; cf. Eq.~\eqref{eq:res1}.

\subsection{Discussion}
	Let us discuss our above findings in some detail.
	We studied the behavior of the CF and its relation to the DNC for states that are defined in finite Fock spaces.
	A one-to-one relation between the degree of the polynomial which defines the CF and the DNC was established.
	Moreover, as an example, we investigated the impact of the photon-addition process on the DNC in such a case.

	The DNC of states of type~\eqref{eq:sup_Fock} is equivalent to the DNC of the highest contributing Fock state. 
	Let us consider the example
	\begin{equation}\label{eq:example_ref}
		|\psi_{m}\rangle=\frac{|0\rangle+\lambda_m|m\rangle}{(1+|\lambda_m|^2)^{1/2}},
	\end{equation}
	with $\lambda_m\in\mathbb{C}$ and $m>0$.
	For $\lambda_m=0$, the state is the classical vacuum state.
	For any, even arbitrarily small, nonzero~$\lambda_m$, this state is nonclassical: $D_{\rm ncl}(|\psi_m\rangle)={m+1}$.
	As the DNC reveals whether or not the state belongs to a set $\mathcal{S}_{n}$ ($1\leqslant n\leqslant m+1$), the jump from $1$ to $m+1$---without attaining intermediate values---is meaningful.
	It demonstrates a direct quantum transition between the nested sets $\mathcal{S}_{1}$ and $\mathcal{S}_{m+1}$ ($\mathcal{S}_{1}\subset\mathcal{S}_{m+1}$) that correspond to no superposition and $m$ superpositions of coherent states, respectively.
	It is also worth mentioning that $|\lambda_m|\to\infty$ for the state $|\psi_{m}\rangle$ results in the Fock state $|m\rangle$.

	A similar discontinuity of the nonclassical depth $\tau_m$~\cite{Lee1991} with respect to changes of the density matrix was also reported in Ref.~\cite{Luetkenhaus1995}.
	For the state~\eqref{eq:example_ref} with $\lambda_m=0$, the nonclassical depth attains the minimal value of $\tau_m=0$ and it discontinuously jumps to become maximal, $\tau_m=1$, for any other value of $\lambda_m\neq0$; cf. Ref.~\cite{Luetkenhaus1995}.
	Note that for applications to measured data, such a discontinuity is less relevant as the change of the state under study has to be significant with respect to error bars of the recorded quantities.
	Further on, all quantum states with a finite number-state expansion (except the vacuum state) have the maximal nonclassicality depth $\tau_m=1$ and are, therefore, indistinguishable by this nonclassicality measure.
	In contrast, the DNC allows for discriminating two states with different, maximal photon numbers.

	For instance, Hillery's distance-based measure \cite{Hillery1987a} also clearly distinguishes different Fock states $|n\rangle$.
	Consistently with the DNC, it becomes stronger for larger $n$~values~\cite{Nair2017,Vogel2014}.
	However, this trace-norm distance between two states has also some limitations for quantifying nonclassicality.
	We can consider two states $|\psi_{m}\rangle$ and $|\psi_{m'}\rangle$ of the type~\eqref{eq:example_ref} which have the same $\lambda_m=\lambda_{m'}\neq0$, but different Fock-state contributions, $0<m<m'$.
	Both states have the same trace-norm distance to the vacuum state $|0\rangle$.
	This leads to the surprising observation that for $\lambda_{m}$ and $\lambda_{m'}$ close to zero, both states have almost the same distance to their closest classical analog.
	Still, to this date, the trace distances could not determined exactly for the states $|\psi_{m}\rangle$ and $|\psi_{m'}\rangle$.
	Only bounds have been given~\cite{Nair2017} for similar states.
	The reason for this issue is that the determination of the closest classical state needs the optimization over an infinite number of variables~\cite{Asboth2005}.
	By contrast, their DNCs are directly accessibly by our method.
	
        The algebraic DNC discerns the amount of nonclassicality of the states $|\psi_m\rangle$ and $|\psi_{m'}\rangle$.
        Hence, it is more sensitive to details of the nonclassicality than the nonclassical depth and more accessible than a distance measure.
        Note that in Ref.~\cite{Sperling2015} a general discussion can be found explaining why certain measures lead to difficulties for quantifying nonclassicality.
	In conclusion, the DNC combines capabilities of the nonclassical depth and the trace distance measure together with a sensitive characterization of nonclassicality.

	Beyond that, the CF can be directly sampled from the experimental data in balanced homodyne detection~\cite{Lvovsky2002,Kiesel2009,Ryl2015}.
	The sampling error grows with $\exp(|\beta|^2/2)$~\cite{Kiesel2009} which is faster than for any polynomial CF.
	Thus, the region where the CF is significant is limited.
	In this region, the CF might be approximated by a truncated Taylor series, which has not necessarily the same polynomial order as the entire CF.
	The experimental effort of recording the CF does not depend on the state under study and, in particular, does not depend on the maximal photon number in Eq.~\eqref{eq:mixFock}.
	In contrast, the number of entries in the density matrix in Fock basis scales quadratically with the maximal photon number of the realized state on top of the above sampling error~\cite{Leonhardt1995}.
	The numerical effort of a state reconstruction is expected to scale correspondingly.
	This is a clear practical advantage of the CF method compared with the distance-based measures.
	
	For theoretical investigations of states with a finite Fock expansion, our above approach is a quite useful tool which describes the fundamental relation of states, which are completely described by finite photon number states, to their amount of nonclassicality.
	However, the experimental applicability might be limited due to the above-discussed growth rate of sampling errors.
	Hence, let us formulate an additional technique to determine the DNC from CFs.

\section{Continuous-Variable Characteristic Functions}\label{sec:Boch}
	In the previous section, we discussed the case of polynomial CFs belonging to finite-dimensional systems.
	An infinite-dimensional system can produce states with CFs exceeding the polynomial behavior.
	In this section, we study such arbitrary CFs.
	First, we recall Bochner's theorem~\cite{Bochner1933} for classical statistics.
	The violation of the corresponding conditions is interpreted in terms of the DNC; i.e., nonclassical states have a DNC larger than~1 (cf.~\cite{Vogel2000}).
	Later on, the lowest-order condition is modified for higher DNC.
	We explicitly present our results for DNC $r=2,\dots,5$ and apply them to some examples.
	The method presented in this section is a witnessing-type approach.
	That is, it gives experiment-friendly criteria to ensure a certain amount of nonclassicality.

\subsection{DNC from the growth of the CF}
	Let us first recall Bochner's theorem in the form of its first application in quantum optics~\cite{Vogel2000,Richter2002}.
	The CF $\Phi(\beta)$ is the Fourier transform of a classical probability density $P_{\rm cl}(\alpha)$ if and only if it satisfies the following properties:
	\begin{itemize}
		\item[(i)] It is normalized [$\Phi(0)=1$].
		\item[(ii)] It is Hermitian [$\forall\beta\in\mathbb{C}{:}$ $\Phi(-\beta)=\Phi(\beta)^{\ast}$].
		\item[(iii)] For any non-negative integer $N$ and any set $\{\beta_{1},\dots,\beta_{N}\}$ of complex numbers, the matrix $[\Phi(\beta_{i}-\beta_{j})]_{i,j=1}^{N}$ is positive semidefinite.
	\end{itemize}
	Conditions (i) and (ii) respectively correspond to the facts that the $P$ function for any quantum state is normalized and real valued~\cite{Richter2002,Vogel2000}, which is true for any state.
	In contrast, constraint (iii) only holds for classical states, i.e., such with a non-negative $P$ function.
	This last criterion allows for a necessary and sufficient identification of nonclassical quantum states.

	In the following, we refer to the dimension $N$ of the matrix $[\Phi(\beta_{i}-\beta_{j})]_{i,j=1}^{N}$ as the order of the classicality condition (iii).
	For instance, the violation of the second-order criterion reads as follows:
	If there exists a $\beta'$ with
	\begin{equation}
		|\Phi(\beta')|>1, \label{eq:Boch2ndV}
	\end{equation}
	then the state $\hat\rho$ is nonclassical~\cite{Vogel2000}.
	Note that this  is a sufficient condition.
	Based on the findings in Ref.~\cite{Hillery1985}, it was shown in Ref.~\cite{KieselThesis} that the nonclassicality condition~\eqref{eq:Boch2ndV} is necessary and sufficient for pure states.
	We can also read this condition in terms of the DNC as follows:
	If the absolute values of the CF of a state $\hat{\rho}$ exceeds the value at the origin, $\Phi(0)=1$, then the DNC of this state exceeds the value of~1:
	\begin{equation}
		\exists\beta':|\Phi(\beta')|>\Phi(0)\Rightarrow D_{\rm ncl}(\hat{\rho})>1. \label{eq:Bochner2nd}
	\end{equation}

\begin{figure}[ht]
	\includegraphics{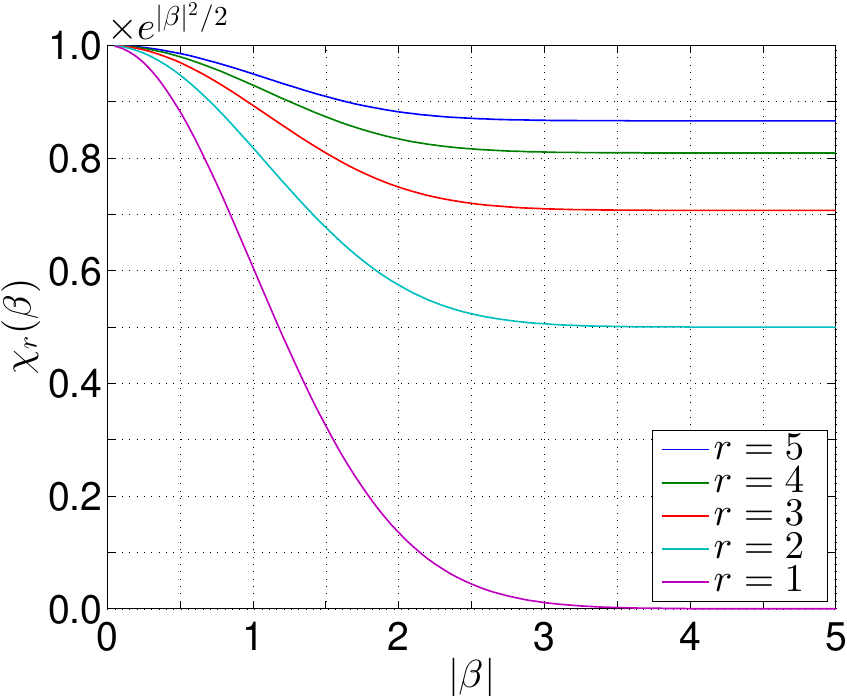}
	\caption{(color online)
		The maximal values $\chi_r(\beta)$ of the CF for different DNCs (from bottom to top: $r=1,\dots,5$).
		For clarity, we have normalized the graph to the maximally possible growth $\exp(|\beta|^2/2)$ of the CF.
		}\label{fig:fracPhi}
\end{figure}

	Now we generalize condition~\eqref{eq:Bochner2nd} to higher DNCs.
	In order to do so, we need to find the maximal modulus of the CF over the set $\mathcal{S}_{r}$ for each $r$ and any~$\beta$.
	In other words, we determine the values of the functions
	\begin{equation}
		\chi_r(\beta)=\sup_{|\Psi_r\rangle\in\mathcal{S}_{r}}|\Phi_{|\Psi_r\rangle}(\beta)|,\label{eq:chir}
	\end{equation}
	where $\Phi_{|\Psi_r\rangle}$ defines the CF of the particular state $|\Psi_r\rangle$ [Eq.~\eqref{eq:Psir}], for $\beta\in\mathbb C$ and $r>1$.
	It is important to mention that the convex structure of states with a DNC of $r$ [cf. Eq.~\eqref{eq:rmix}] implies that the bound $\chi_r$ also applies to all mixed states in $\mathcal S_r$.
	Clearly, from condition~\eqref{eq:Bochner2nd}, we get $\chi_1(\beta)=1$ for $r=1$.
	For $r=2,\dots,5$, we calculated the bounds $\chi_r$ numerically (see the Appendix), which can be straightforwardly extended to $r>5$.
	The resulting functions $\chi_r$ are shown in Fig.~\ref{fig:fracPhi}.
	Details of the plot are discussed in the next section.

	Let us conclude our findings.
	The CF conditions for identifying a DNC read as follows:
	\begin{corollary}
		If the absolute value of the CF $\Phi$ of the state $\hat{\rho}$ exceeds the depicted values in Fig.~\ref{fig:fracPhi} of $\chi_{r}$ at some point $\beta'$, the DNC of this state $\hat{\rho}$ is larger than~$r$; equivalently,
		\begin{equation}
			\exists\beta': |\Phi(\beta')|>\chi_{r}(\beta')\Rightarrow D_{\rm ncl}(\hat{\rho})>r.\label{eq:criteria}
		\end{equation}
	\end{corollary}
	In the continuation of this section, we further analyze our results and, eventually, apply them to some examples.
	
\subsection{Numerical results}
	Let us discuss Fig.~\ref{fig:fracPhi}.
	At the origin, $\beta=0$, we have $\chi_r(0)=1$ as for any CF holds $\Phi(0)=1$ [condition (i)].
	The different curvatures of $\chi_r$ around the origin ($|\beta|\approx0$) can be explained with the squeezed quadrature variances shown by states of type~\eqref{eq:Psir}; see also Ref.~\cite{Mraz2014}.
	Around $\beta\approx3.5$, the values of the normalized $\chi_r$ tend to be a constant fraction of the CF of the maximally singular $P$ function~\cite{Sperling2016}, which is given by $\Phi(\beta)\exp(-|\beta|^2/2)=1$.
	Even the value of the plateaus in Fig.~\ref{fig:fracPhi} can be explained by having a closer look at the positions of the supremum~\eqref{eq:chir}, for which the details can be found in the Appendix.
	The asymptotic values of the plateaus in Fig.~\ref{fig:fracPhi} are listed in Table~\ref{tab:plateaus}.
	In the limit $r\to\infty$, the bound is $\chi_\infty(\beta)=\exp(|\beta|^2/2)$, as the general bound of the CF for arbitrary states is this function and this bound can be approached by physical states~\cite{Sperling2016}.
	In other words, the values of the plateaus in Fig.~\ref{fig:fracPhi} converge to 1 for $r\to\infty$.
	Furthermore, the bounds $\chi_r$ solely depend on $|\beta|$ as a rotation in phase space (or in the Fourier-transformed space) is a classical operation.
	Analogously to the presented approach, conditions for the DNC can be constructed from other nonclassicality conditions; see, e.g., Refs.~\cite{Richter2002,Ryl2015}.

\begin{table}[ht]
	\caption{
		The asymptotic values of plateaus in Fig.~\ref{fig:fracPhi}, i.e., the values $\lim_{|\beta|\to\infty}\chi_r(\beta)e^{-|\beta|^2/2}$, are listed for different DNCs ($r$).
		}\label{tab:plateaus}
	\begin{tabular}{cc}
		\hline\hline
		$r$&Plateau\\\hline
		1 & 0\\
		2 & $1/2$\\
		3 & $1/\sqrt{2}\approx0.7071$\\
		4 & $(1+\sqrt{5})/4\approx0.8090$\\
		5 & $\sqrt{3}/2\approx0.8660$ \\
		$\vdots$ & $\vdots$ \\
		$\infty$ & $1$ \\
		\hline\hline
	\end{tabular}
\end{table}
	
\subsection{Example: squeezed state}\label{sec:Examples}
	To apply our method, let us start with properties of a pure squeezed vacuum state,
	\begin{equation}
		|\xi;0\rangle=\sum_{n=0}^{\infty}\frac{(e^{i\arg(\xi)}\tanh|\xi|)^n}{2^nn!\sqrt{\cosh|\xi|}}\sqrt{(2n)!}|2n\rangle.\label{eq:sq_th}
	\end{equation}
	Its DNC is known~\cite{Vogel2014} to be infinite in theory for any $\xi\neq0$.
	However, it becomes harder to witness this infinite DNC for decreasing squeezing strengths $|\xi|$ \cite{Mraz2014} and is discussed below.
	For this state, the minimal (squeezed) and maximal (antisqueezed) variances of the quadrature operator 
	\begin{equation}
		\hat{x}(\varphi)=e^{-i\varphi}\hat{a}^\dag+e^{i\varphi}\hat{a} \label{eq:quadrOp}
	\end{equation}
	can be given in terms of the squeezing strength as
	\begin{align}
		V_{\rm sq}=\min\limits_\varphi\langle [\Delta\hat{x}(\varphi)]^2\rangle =e^{-2|\xi|},\nonumber\\ V_{\rm asq}=\max\limits_{\varphi}\langle [\Delta\hat{x}(\varphi)]^2\rangle =e^{2|\xi|}.
	\end{align}
	Such pure squeezed states are also called minimal uncertainty states since $V_{\rm sq}V_{\rm asq}=1$.
	It is also common to specify the squeezing $S$ in decibels (dB):
	\begin{equation}
		S=10\log_{10}\frac{\min\limits_\varphi\langle [\Delta\hat{x}(\varphi)]^2\rangle}{\langle[\Delta\hat{x}(0)]^2\rangle_{\rm vac}}=-\frac{20|\xi|}{\ln(10)}.
	\end{equation}
	For vacuum holds $S=0$.

	The DNC is independent of the value of the squeezing parameter $\xi$.
	At first sight, this may be counterintuitive as other measures like the nonclassical depth are $|\xi|$ dependent~\cite{Lee1991}.
	Equation~\eqref{eq:DNC_sq} holds for any $\xi\neq0$, i.e., as long as all terms of the series~\eqref{eq:sq_th} contribute to the state.
	The weight factors of Fock states in the series~\eqref{eq:sq_th} converge to zero for $n\to\infty$, which is necessary for the series to be convergent.
	Any measured state has unavoidably a finite statistical significance so that only a finite number of summands in~\eqref{eq:sq_th} contributes significantly to the measured data.
	Such a truncated state is statistically indistinguishable from an infinite superposition of number states.
	Thus, only a finite DNC can be certified.
	For a weakly squeezed state, $|\xi|\ll1$, the truncated state is approximately 
	\begin{equation}\label{eq:sq_approx}
		|\xi;0\rangle\approx\frac{\sqrt{2}|0\rangle+\xi|2\rangle}{\sqrt{2+|\xi|^2}},
	\end{equation}
	which is of type~\eqref{eq:example_ref}.
	Therefore, the maximally certifiable amount of nonclassicality is also bounded.
	In the case of the here-studied witnessing-based estimation of the DNC, the limit is due to the statistical significance.
	Therefore, a degree of nonclassicality of at least $r$ is ensured as long as the violation of the bounds $\chi_r(\beta)$ is significant.
	Also note that such states are used to generate photon pairs, for more details, see, for instance, Refs.~\cite{Walmsley2008,Smith2013}.
	Thus, from the experimental point of view, the verifiable DNC indeed depends on the squeezing strength; see also Ref.~\cite{Mraz2014}.

\begin{figure}[ht]
	\includegraphics{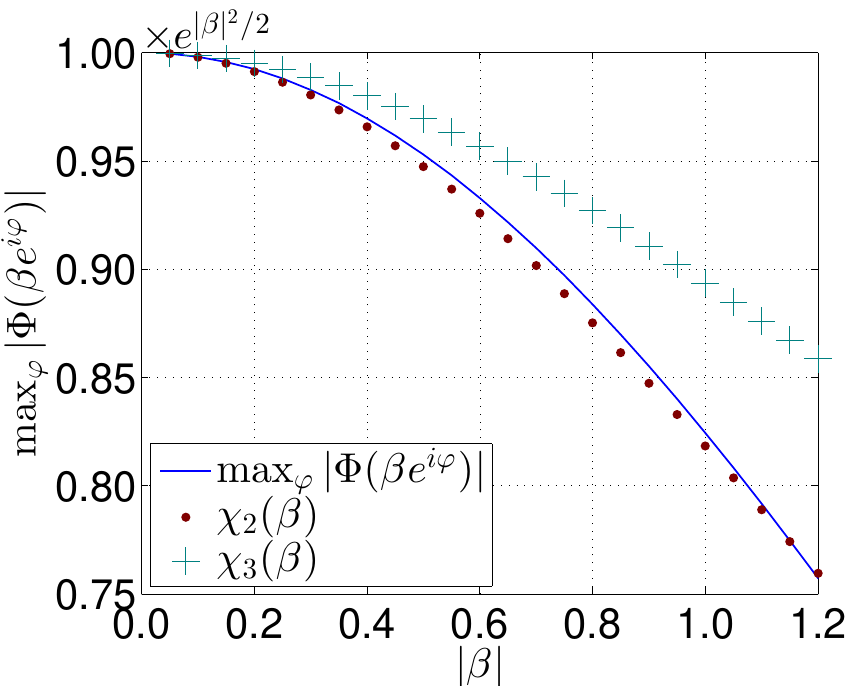}
	\caption{(color online)
		For the state $\hat{\rho}_{\rm ex}$ from Ref.~\cite{Ryl2015}, the squeezed (solid curve) direction of the CF is shown, $\max_\varphi |\Phi(\beta e^{i\varphi})|$.
		It exceeds the boundary $\chi_2(\beta)$ (dots), proving the DNC of this state to be at least three.
		Moreover, it is upper bounded by $\chi_3(\beta)$ (crosses).
		}\label{fig:example1}
\end{figure}

	Beyond pure states, we use a mixed and moderately squeezed vacuum state $\hat{\rho}_{\rm ex}$ to demonstrate the applicability of our conditions~\eqref{eq:criteria}.
	Such a state was experimentally characterized in Ref.~\cite{Ryl2015}; it exhibits $S_{\rm ex}=-4.13\,{\rm dB}$ of squeezing and $+6.11\,{\rm dB}$ of antisqueezing.
	Since the antisqueezing is stronger than the squeezing, the state is not pure, but it includes some classical noise.
	For the state $\hat{\rho}_{\rm ex}$, we can prove $D_{\rm ncl}(\hat{\rho}_{\rm ex})\geqslant3$; see Fig.~\ref{fig:example1}.
	It is worth pointing out that this result is in agreement with the boundaries for the squeezing strength given in Ref.~\cite{Mraz2014}: 
	\begin{equation}
		S_3=-5.91\,{\rm dB}<S_{\rm ex}=-4.13\,{\rm dB}<-3.54\,{\rm dB}=S_2.
	\end{equation}
	
\subsection{Example: squeezed state with vacuum}
\begin{figure}[ht]
	\includegraphics{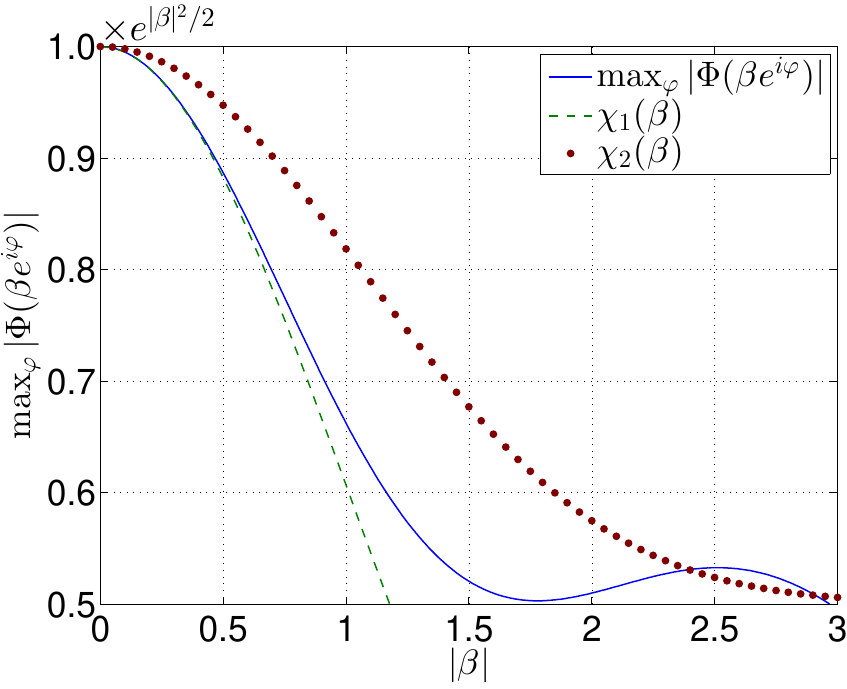}
	\caption{(color online)
		For $\xi=-0.562$ (or $S=-4.88\,{\rm dB}$) and $\lambda=-1.4$ the state \eqref{eq:example} shows no squeezing~\eqref{eq:example_prop} (solid curve).
		The curve for $|\beta|\ll1$ is even below the classical limit $\chi_1(\beta)$.
		However, our CF criterion proves the DNC of this state to be bigger than two since its CF even exceeds the boundary $\chi_2(\beta)$ for larger $|\beta|$ values.%\in(2.4;2.9)$.
	}\label{fig:example2}
\end{figure}

	Now let us show how quantum superpositions can influence the quadrature variances and the quantification of nonclassicality.
	For this reason, we consider a second example which is a superposition of vacuum state and squeezed vacuum state,
	\begin{equation}
		|\psi\rangle=\mathcal{N}(|0\rangle+\lambda|\xi;0\rangle),\label{eq:example}
	\end{equation}
	with $\xi$ being the complex-valued squeezing parameter and $\mathcal{N}$ the normalization constant.
	Its CF reads
	\begin{eqnarray}
		\Phi(\beta)
		&=&\mathcal{N}^2\left(1+\frac{\lambda^{\ast}}{\sqrt{\mu}} e^{-\beta^{2}\nu^{\ast}/(2\mu)}+\frac{\lambda}{\sqrt{\mu}}e^{-\beta^{\ast 2}\nu/(2\mu)}\right.\nonumber\\
		&&\left.\phantom{\frac{0}{1}}+|\lambda|^2e^{|\beta|^2/2-|\mu\beta+\nu\beta^{\ast}|^2/2}\right).
	\end{eqnarray}
	Here we used the typical parametrization where $\mu=\cosh|\xi|$ and $\nu=e^{i\arg\xi}\sinh|\xi|$.
	The normalization constant is $\mathcal{N}^2=[1+(\lambda+\lambda^{\ast})/\sqrt{\mu}+|\lambda|^2]^{-1}$.
	For specific parameters, e.g., $\xi=-0.562$ and $\lambda=-1.4$, the state $|\psi\rangle$ shows quadrature variances above the vacuum level for all phases:
	\begin{equation}
		\langle[\Delta \hat{x}(\varphi)]^2\rangle>1\,\forall\varphi
		\text{ or }
		\min\limits_{\varphi}\langle[\Delta \hat{x}(\varphi)]^2\rangle\approx1.0006. \label{eq:example_prop}
	\end{equation}
	So the previously considered squeezing conditions~\cite{Mraz2014} cannot identify the nonclassicality of the state.
	Figure~\ref{fig:example2} displays the maximally growing direction of the CF ($\max_\varphi |\Phi(\beta e^{i\varphi})|$) of the state~\eqref{eq:example}.
	It shows no violation of classicality  for $|\beta|{\ll}1$.
	For larger values of $\beta$ the CF exceeds $\chi_2(\beta)$, which directly proves that
	\begin{equation}
		D_{\rm ncl}(|\psi\rangle)\geqslant 3.
	\end{equation}
	Our criteria in terms of CFs use more information than just the squeezing level.
	Thus, they can outperform the previously considered quadrature-variance-based conditions.

	Another way to combine states is a classical mixture, here, of vacuum and squeezed vacuum states,
 	\begin{equation}
		\hat{\rho}=\frac{1}{1+|\lambda|^2}(|0\rangle\langle0|+|\lambda|^2|\xi;0\rangle\langle\xi;0|).\label{eq:example2}
 	\end{equation}
	The CF of such mixtures reads as
	\begin{equation}
		\Phi(\beta)=\frac{1}{1+|\lambda|^2}\left(1+|\lambda|^2e^{|\beta|^2/2-|\mu\beta+\nu\beta^{\ast}|^2/2}\right).
	\end{equation}
	For $\lambda=0$, the mixture becomes a pure vacuum state and for $\lambda\to\infty$ a pure squeezed state.
	By definition, the state depends only on the absolute value of $\lambda$ and not on its phase.
	In comparison to the state in Eq.~\eqref{eq:example}, the state~\eqref{eq:example2} has no quantum interferences between the classical vacuum state and the squeezed one.
	
\begin{figure}[ht]
	 \includegraphics{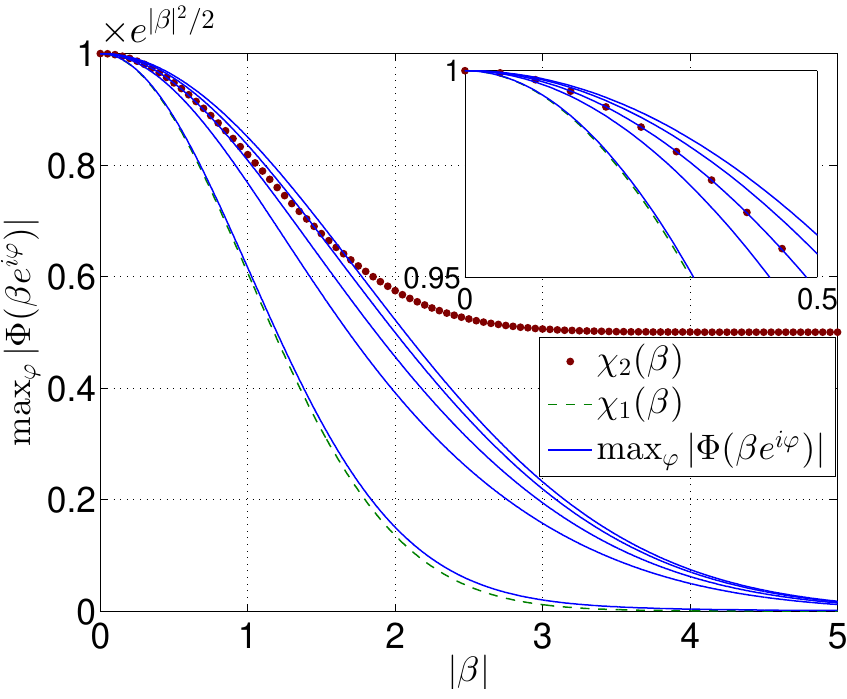}
	 \caption{(color online)
		  For the squeezing parameter $\xi=-0.562$ and $|\lambda|=0.2,1.4,2.2,4,\infty$ (solid curves from left to right), the CFs of the mixture~\eqref{eq:example2} are shown.
		  For $0<|\lambda|<2.2$, a DNC of at least two can be proven, as the limit $\chi_1(\beta)$ (dashed line) is exceeded.
		  For $|\lambda|\gtrsim 2.2$, also a DNC of at least three can be verified since $\chi_2(\beta)$ (dotted line) is surpassed.
		  The inset magnifies the region close to the origin, $\beta=0$.}\label{fig:examplemix}
\end{figure}

	Figure~\ref{fig:examplemix} shows the maximally growing direction of the CF [$\max_\varphi |\Phi(\beta e^{i\varphi})|$] of the states~\eqref{eq:example2} for different weight factors $\lambda$.
	The verifiable DNC of these states depends on the value of this weight factor.
	We used the same squeezing parameter $\xi=-0.562$ as in Fig.~\ref{fig:example2}.
	For $\lambda$ close to zero ($\hat\rho$ close to vacuum state), the CF exceeds only the boundary $\chi_1(\beta)$.
	For $\lambda=-1.4$, the CF shows a different behavior to that of state~\eqref{eq:example}.
	It exceeds only $\chi_1(\beta)$ and not $\chi_2(\beta)$ (dotted line).
	So, only $D_{\rm ncl}(\hat\rho)\geqslant 2$ can be verified.
	The value $|\lambda|\approx2.2$ is roughly the threshold for proving $D_{\rm ncl}(\hat\rho)\geqslant3$ by crossing $\chi_2(\beta)$. 
	For the mixture with $\lambda=4$ and for the pure squeezed state ($\lambda=\infty$), $D_{\rm ncl}(\hat\rho)\geqslant 3$ is proven.
	With decreasing contribution of the squeezed state, i.e., for decreasing $|\lambda|$, the certified DNC also decreases.
	In fact for $\lambda\to0$, the state~\eqref{eq:example2} converges to the vacuum state and detecting a DNC larger than one becomes more and more difficult.
	Note that we have a DNC of one for the vacuum state itself.

	Comparing both examples, we can see that quantum superpositions can, similarly to classical mixtures, reduce the quadrature variance.
	However, the case of a superposition also has the capability to preserve or---for other types of states---even increase the DNC.
	Conversely, the mixture with a classical state is a classical operation and therefore does not have this potential.
	The contribution of the superposition terms allows us to prove higher DNC for state~\eqref{eq:example} compared with state~\eqref{eq:example2} for the considered parameters $\xi$ and $\lambda$.
	This comparison of the superposition and the mixture of a squeezed state with vacuum also highlights the general impact of quantum superpositions, which we use for our quantification technique.

\section{Summary and Conclusions}\label{sec:Conc}
	In summary, we studied the relation between the characteristic function and the amount of nonclassicality.
	Two approaches have been pursued.
	First, a complete analysis of polynomial characteristic functions has been performed analytically.
	We have shown that they always correspond to states which have a finite Fock-state expansion.
	Most importantly, the degree of the polynomial directly connects to the degree of nonclassicality of the state under study.
	In this context, we also investigated the impact of photon-addition protocols on the amount of nonclassicality in such a case.
	
	Our second approach applies to arbitrary characteristic functions.
	We numerically computed the bounds of the characteristic function with a given degree of nonclassicality.
	If these bounds are exceeded by experimentally measured characteristic functions, then this state has the degree of nonclassicality which is larger than given by that bound.
	We applied this criterion to squeezed states and discussed the relation to a previously considered quadrature-witness approach.
	For a superposition of vacuum with a squeezed state, which does not show squeezing, we could even show that our condition is stronger than the quadrature-based method.
	As the characteristic function can be directly sampled from balanced-homodyne-detection data, our criteria present a directly applicable approach to quantify the amount of nonclassicality of light fields generated in experiments.
	It is also worth pointing out that the degree of nonclassicality can be directly mapped onto the amount of entanglement realized in the output ports of a passive optical device~\cite{Vogel2014}.

	Our chosen quantifier---the degree of nonclassicality---was discussed in relation to other measures and its deeper connection to the quantum superposition principle was highlighted.
	We explained how discontinuities link to transitions between different levels of quantum superpositions to enlarged alphabets for information encoding.
	While our first approach results in an exact analysis of the degree of nonclassicality, the second, measurement-friendly technique certifies the usable amount of nonclassicality of an experimentally realized state.
	
\begin{acknowledgments}
	The authors are grateful to B.~K\"uhn and D.~Yu.~Vasylyev for helpful discussions.
	This work has been supported by the Deutsche Forschungs\-gemeinschaft through SFB~652 (project B2).
	J.~S. and W.~V. acknowledge the project leading to this application which has received funding from the European Union's Horizon 2020 research and innovation program Grant No. 665148.
\end{acknowledgments}

\appendix*

\section{Numerical solution and additional discussions}\label{appNum}
	First, let us discuss the case $r=2$.
	General superpositions,
	\begin{equation}
		|\Psi_{2}\rangle=\mathcal{N}(\lambda_{1}|\alpha_{1}\rangle+\lambda_{2}|\alpha_{2}\rangle,\label{eq:C1}
	\end{equation}
	are parametrized by eight real-valued (four complex-valued) variables ($\lambda_1,\lambda_2,\alpha_1,\alpha_2\in\mathbb C$).
	As the global phase plays no role, there are still seven real variables.
	Hence, we need to find the optimum over the set $\mathbb{R}^7$.
	Since the DNC is conserved under displacement~\cite{Vogel2014}, we can set one of the coherent states to vacuum, $\alpha_{1}=0$, and the other coherent amplitude to $\alpha_{2}-\alpha_{1}=\alpha$.
	Via a rescaling of the normalization constant $\mathcal{N}$, we can choose $\lambda_{1}=1,\lambda_{2}/\lambda_{1}=\lambda$.
	Now only four real variables are left for the optimization,
	\begin{equation}
		|\Psi_{2}\rangle=\mathcal{N}(|0\rangle+\lambda|\alpha\rangle),\label{eq:C2}
	\end{equation}
	Such transformations allow us to simplify the dimension of the optimization problem from $\mathbb{R}^8$ to $\mathbb{R}^4$.
	Analogously, for DNC $r$, the number of (real-valued) variables can be reduced from $4r$ to $4(r-1)$ for reducing the complexity of the numerics.

 	The optimization over the states~\eqref{eq:Psir},
 	\begin{equation}
		\chi_{r}(\beta)=\max\limits_{\substack{[{\rm Re}(\lambda_2),\dots,{\rm Im}(\lambda_r),\qquad\qquad\\{\rm Re}(\alpha_2),\dots,{\rm Im}(\alpha_r)]^{\rm T}\in\mathbb{R}^{4(r-1)}}}|\Phi_{|\Psi_r\rangle}(\beta)|,
 	\end{equation}
 	was performed by an iterative search routine that determines the root of the gradient by the gradient ascent method.
 	The starting points for the iteration were set randomly and multiple runs have been executed to ensure the stability of our technique.
 	We used about $2^{4(r-1)}$ starting points for each value of $\beta$ to ensure roughly one starting point per signs configuration of the real-valued parameters.
	Based on this proceeding, we determined the desired values of $\chi_{r}(\beta)$.

\begin{figure}[ht]
	 \includegraphics{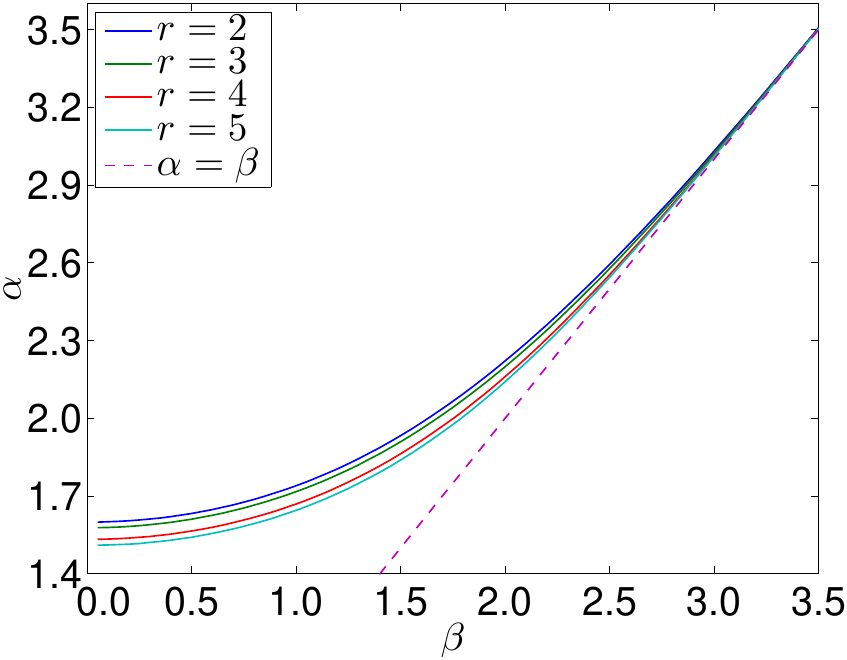}
	 \caption{(color online)
		The position $\alpha$ of the maximum of the CF for the DNC from top to bottom: $r=2,\dots,5$.
		The dashed line shows the asymptote $\alpha=\beta$.
	}\label{fig:alphas}
\end{figure}

	For each $r$ and specified value of $\beta$, the states of type~\eqref{eq:Psir}, leading to the maximal value of the CF, were calculated; i.e., the values $(\lambda_2,\dots,\lambda_r,\alpha_2,\dots,\alpha_r)$ were determined.
	The formal structure of these states is given in Table~\ref{tab:states}.
	The weight factors $\lambda_i$ were nearly constant in the investigated interval $|\beta|\in[0.05,5]$.
	The optimal coherent amplitudes $\alpha_i$ have been found to reveal a specific, interesting structure(Table~\ref{tab:states}).
	Further, they depend on $\beta$ as shown in Fig.~\ref{fig:alphas}.

\begin{table}[ht]
	\caption{
		The structure of the states that produce the highest values of CF is listed for $r=2,\dots,5$.
		For the slightly varying weight factors, the closed-form approximations reproducing the values of the plateaus in Fig.~\ref{fig:fracPhi} are given.
		The coherent amplitudes are equidistantly distributed but depend on $\beta$ as shown in Fig.~\ref{fig:alphas}.
	}\label{tab:states}
	\begin{tabular}{cc l}
		\hline\hline
		DNC && State\\
		\hline
		$r=2$&&$|\Psi_2\rangle=\mathcal{N}(|0\rangle+|\alpha\rangle)$\\
		$r=3$&&$|\Psi_3\rangle=\mathcal{N}(|0\rangle+\sqrt{2}|\alpha\rangle+|2\alpha\rangle)$\\
		$r=4$&&$|\Psi_4\rangle=\mathcal{N}(|0\rangle+\frac{1+\sqrt{5}}{2}|\alpha\rangle+\frac{1+\sqrt{5}}{2}|2\alpha\rangle+|3\alpha\rangle)$\\
		$r=5$&&$|\Psi_5\rangle=\mathcal{N}(|0\rangle+\sqrt{3}|\alpha\rangle+2|2\alpha\rangle+\sqrt{3}|3\alpha\rangle+|4\alpha\rangle)$\\
		\hline\hline
	\end{tabular}
\end{table}

	From Fig.~\ref{fig:alphas}, we can see that there are two asymptotic regions.
	For $\beta\to0$, the value of $\alpha$ does not change strongly.
	Thus, in this region, the state $|\Psi_r\rangle$ is nearly constant.
	As described later on, the resulting $\chi_r(\beta)$ can be explained with quadrature variances below vacuum level which belong to the states $|\Psi_r\rangle$.
	The other asymptotic regime which is discussed is $\beta\gg1$, which is represented by the curve $\alpha=\beta$.

\subsubsection*{Case $|\beta|\ll1$}
	For $|\beta|$ close to zero, the bounds do not change strongly as shown in Fig.~\ref{fig:alphas}.
	Therefore, we can approximate $\chi_r(\beta)$ with a finite Taylor expansion.
	The same holds true for the CF of the state that yields this bound, $\chi_r(\beta)=|\Phi(\beta)|$.
	That is, the CF can be described by the Taylor series around the point $\beta{=}0$,
	\begin{equation}
		\Phi(\beta)=\sum\limits_{k,l=0}^{\infty} \partial_{\beta}^k\partial_{\beta^\ast}^l\Phi(\beta)\Big|_{\beta=0}\frac{\beta^k\beta^{\ast l}}{k!l!}.
	\end{equation}
	The values of the derivatives at $\beta{=}0$ are connected  with the moments of the $P$ function,
	\begin{equation}
		\partial_{\beta}^k\partial_{\beta^\ast}^l\Phi(\beta)\Big|_{\beta=0}=(-1)^l\langle\hat{a}^{\dag k}\hat{a}^l\rangle
	\end{equation}
	(see also Sec.~\ref{subsec:PolyCF}).
	For $|\beta|\ll1$, the functions $\chi_r(\beta)=|\Phi(\beta)|$, using the CF of the optimal state, can be approximated by low-order Taylor polynomials:
	\begin{eqnarray}
		\Phi(\beta)&=&1{-}\langle\hat{a}\rangle\beta^\ast+\langle\hat{a}^\dag\rangle\beta\nonumber\\
		&&+\frac{\langle\hat{a}^{\dag 2}\rangle}{2}\beta^2-\langle\hat{a}^\dag\hat{a}\rangle|\beta|^2+\frac{\langle\hat{a}^2\rangle}{2}\beta^{\ast 2}+O(|\beta|^3)\nonumber\\
		&=&1+i|\beta|\langle\hat{x}[\arg(-i\beta)]\rangle\nonumber\\
		&&+\frac{(i|\beta|)^2}{2}\langle{:}\{\hat{x}[\arg(-i\beta)]\}^2{:}\rangle+O(|\beta|^3).
	\end{eqnarray}
	
	The previously studied witnessing condition for quadrature squeezing needs only moments up to second order~\cite{Mraz2014}.
	There, the quantification was studied in terms of minimal quadrature variances.
	The bounds,
	\begin{equation}
		\inf\limits_{|\Psi_r\rangle\in\mathcal{S}_r}\min\limits_{\varphi}\langle\Psi_r|{:}[\Delta \hat{x}(\varphi)]^2{:}|\Psi_r\rangle=\exp(-2|\xi_r|),
	\end{equation}
	were calculated for the squeezed quadrature variances (or squeezing parameters $\xi_r$); see Ref.~\cite{Mraz2014}.
	The conditions for DNC in terms of squeezed variances reads:
	\begin{equation}
		\min\limits_{\varphi}{\rm Tr}\{\hat{\rho}[\Delta\hat{x}(\varphi)]^2\} <e^{-2|\xi_r|}\Rightarrow D_{\rm ncl}(\hat{\rho})>r.
	\end{equation}

\begin{figure}[ht]
	 \includegraphics{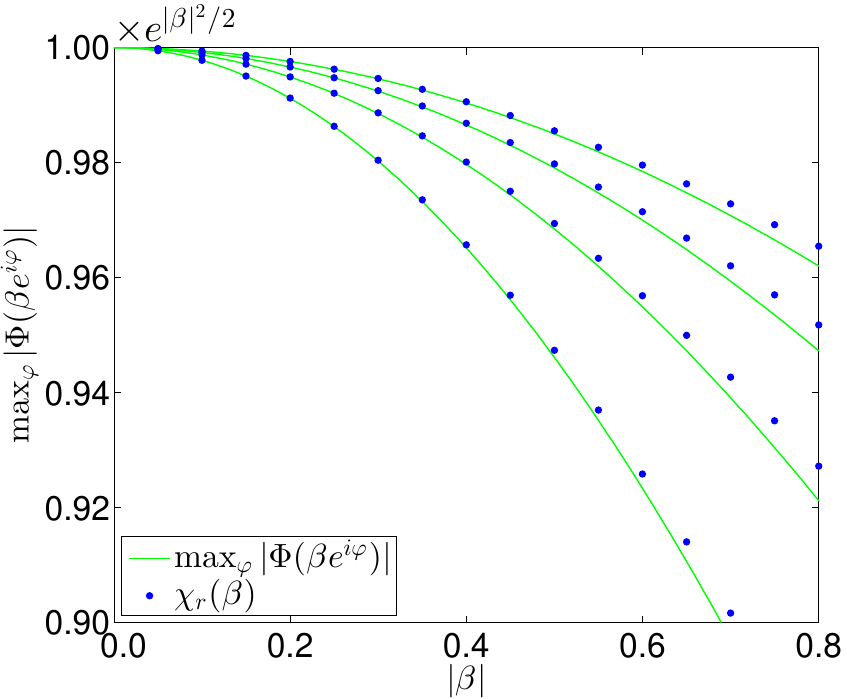}
	\caption{(color online)
		Solid lines are calculated from maximal squeezing parameters $\xi_r$ for DNC $r=2,\dots,5$ (from left to right) given in Ref.~\cite{Mraz2014}.
		For comparison, the dots are the values of $\chi_r(\beta)$.
		For small $|\beta|$, they coincide.
		For larger $|\beta|$ values, higher-order moments start to be relevant and results start to differ.
	}\label{fig:compMraz}
\end{figure}

	For small $|\beta|$, Fig.~\ref{fig:compMraz} shows---as expected in the given second-order approximation---a good agreement of squeezing curves for $\xi_r$ and the dots give the values of $\chi_r(\beta)$.
	However, we clearly observe deviations for $|\beta|\gtrsim0.3$ and the estimated $\chi_r(\beta)$ cannot be solely explained with second-order moments.
	Similarly in Fig.~\ref{fig:alphas}, around the same value $\beta\approx0.3$, the position $\alpha$ of maximum $\chi_r(\beta)$ starts to change with $\beta$.

\subsubsection*{Case $|\beta|\gg1$}
	In order to understand the result for large coherent amplitudes $|\beta|$, in particular the plateaus in Fig.~\ref{fig:fracPhi}, we take a closer look at $\Phi(\beta)$ for a normalized pure state with a DNC of $r$:
	\begin{align}
		&e^{-|\beta|^2/2}\Phi(\beta) \label{eq:Phi_W}
		\\\nonumber=&\frac{
		\sum\limits_{k,l=1}^{r}\lambda_{k}^{\ast}\lambda_{l} e^{\beta\alpha_{k}^{\ast}-\beta^{\ast}\alpha_{l}+\alpha_{k}^{\ast}\alpha_{l}-(|\alpha_{k}|^{2}+|\alpha_{l}|^{2}+|\beta|^2)/2}
		}
		{
		\sum\limits_{k,l=1}^{r}\lambda_{k}^{\ast} \lambda_{l}
		e^{-|\alpha_{k}|^{2}/2-|\alpha_{l}|^{2}/2+\alpha_{k}^{\ast}\alpha_{l}}
		}.
	\end{align}
	The exponent in the numerator has the real part $-|\beta-(\alpha_{k}-\alpha_{l})|^{2}/2$
	and the imaginary part $[\beta(\alpha_{k}+\alpha_{l})^{\ast}-\beta^{\ast}(\alpha_{k}+\alpha_{l})]/2+(\alpha_{k}^{\ast}\alpha_{l}-\alpha_{k}\alpha_{l}^{\ast})/2$.
	The latter changes only the phase of the summands but not their absolute values.

	As the case $r=1$ is trivial, we start with the case $r=2$.
	After the simplifications which are described at the beginning of this Appendix, the CF for the state~\eqref{eq:C2} obeys
	\begin{align}
	\begin{aligned}
		\Phi(\beta)=&e^{|\beta|^2/2}\left[\frac{(1+|\lambda|^2e^{2i\varphi})e^{-|\beta|^2/2}}{1+(\lambda+\lambda^{\ast})e^{-|\alpha|^{2}/2}+|\lambda|^2} \right.\\
		&+\left.\frac{e^{i\varphi}(\lambda^{\ast} e^{-|\alpha-\beta|^{2}/2}+\lambda e^{-|\alpha+\beta|^{2}/2})}{1+(\lambda+\lambda^{\ast})e^{-|\alpha|^{2}/2}+|\lambda|^2}\right],
	\end{aligned}
	\end{align}
	with $\varphi=(\beta\alpha^{\ast}-\beta^{\ast}\alpha)/(2i)$.
	For $|\beta|\gg 1$, the numerator of the first fraction is almost zero, but the numerator of the second one can have larger values.
	For $\alpha=\pm\beta$, one of the terms $\exp(-|\alpha\mp\beta|^2/2)$ becomes~1.
	Let us consider the case $\alpha=\beta$, i.e., $\alpha-\beta=0$ and $\alpha+\beta=2\beta$.
	A slight perturbation of $\alpha$ would significantly influence $\exp(-|\alpha-\beta|^2/2)$, whereas $\exp(-|\alpha+\beta|^2/2)$ is still almost zero.
	If we set $\alpha\approx0$ [$\alpha=0$ is forbidden by definition in Eq.~\eqref{eq:Psir}], both summands become roughly $\exp(-|\beta|^2/2)$; hence, they are close to zero.
	The case $\lambda=-1,\alpha\to0$ corresponds to Fock state $|1\rangle$ {(cf.~\cite{Vogel2014})}.
	Thus, the only way to maximize the sum is setting $\alpha=\pm\beta$, which yields the first plateau in Table~\ref{tab:plateaus}.
	
	For a given $r=r'$, we similarly assume that $\alpha_1,\dots,\alpha_{r'}$ fulfill $\alpha_{k}-\alpha_{k-1}=\beta$ for $k=2,\dots,r'$.
	For $r=r'+1$, we have to set $\alpha_{r'+1}$.
	Again, with similar arguments as in the case $r=2$, $\alpha_{r'+1}=\alpha_{r'}+\beta$ (or $\alpha_{r'+1}=\alpha_1-\beta$) is the optimal choice of the value of $\alpha_{r'+1}$ in Eq.~\eqref{eq:Phi_W}.
	For such a choice, we get $\alpha_{k}-\alpha_{l}=(k-l)\beta$.
	This choice leads to almost orthogonal states $\langle\alpha_{k}|\alpha_{l}\rangle\ll 1$ $(k\neq l)$ for $|\beta|\gg 1$.
	For this reason, the denominator in Eq.~\eqref{eq:Phi_W} can be replaced by the single sum over $k=l$.
	Only a part of the numerator terms is not damped by $\exp(-|\beta|^2/2)$, these with $\alpha_{k}-\alpha_{l}=\beta$.
	The quotient of these constant terms and the denominator leads to the value of the plateaus in Fig.~\ref{fig:fracPhi}.


\begin{thebibliography}{100}
	\bibitem{Mandel1979} L. Mandel, {\it Sub-Poissonian photon statistics in resonance fluorescence}, \href{http://ol.osa.org/abstract.cfm?URI=ol-4-7-205}{Opt. Lett. {\bf 4}, 205 (1979)}.
	\bibitem{Slusher1985} R. E. Slusher, and L. W. Hollberg, and B. Yurke, and J. C. Mertz and J. F. Valley, {\it Observation of squeezed states generated by four-wave mixing in an optical cavity}, \href{http://dx.doi.org/10.1103/PhysRevLett.55.2409}{Phys. Rev. Lett. {\bf 55}, 2409 (1985)}.
	\bibitem{Titulaer1965} U. M. Titulaer and R. J. Glauber, {\it Correlation Functions for Coherent Fields}, \href{http://link.aps.org/doi/10.1103/PhysRev.140.B676}{Phys. Rev. {\bf 140}, B676 (1965)}.
	\bibitem{Mandel1986} L. Mandel, {\it Non-classical states of the electromagnetic field}, \href{http://stacks.iop.org/1402-4896/1986/i=T12/a=005}{Phys. Scr. {\bf T12}, 34 (1986)}.
	\bibitem{Glauber1963} R. J. Glauber, {\it Coherent and incoherent states of the radiation field}, \href{http://dx.doi.org/10.1103/PhysRev.131.2766}{Phys. Rev. {\bf 131}, 2766 (1963)}.
%	
	\bibitem{Sudarshan1963} E. C. G. Sudarshan, {\it Equivalence of semiclassical and quantum mechanical descriptions of statistical light beams}, \href{http://dx.doi.org/10.1103/PhysRevLett.10.277}{Phys. Rev. Lett. {\bf 10}, 277 (1963)}.
	\bibitem{Hillery1987a} M. Hillery, {\it Nonclassical distance in quantum optics}, \href{http://link.aps.org/doi/10.1103/PhysRevA.35.725}{Phys. Rev. A {\bf 35}, 725 (1987)}.
	\bibitem{Asboth2005} J. K. Asb\'{o}th, J. Calsamiglia, and H. Ritsch, {\it Computable Measure of Nonclassicality for Light}, \href{http://link.aps.org/doi/10.1103/PhysRevLett.94.173602}{Phys. Rev. Lett. {\bf 94}, 173602 (2005)}.
	\bibitem{Nair2017} R. Nair, {\it Nonclassical distance in multimode bosonic systems}, \href{https://arxiv.org/abs/1701.07688}{arXiv:1701.07688 [quant-ph]}.
	\bibitem{Wuensche2001} A. W\"{u}nsche, V. V. Dodonov, O. V. Man'ko, and V. I. Man'ko, {\it Nonclassicality of States in Quantum Optics}, \href{http://onlinelibrary.wiley.com/doi/10.1002/1521-3978(200110)49:10/11<1117::AID-PROP1117>3.0.CO;2-4/abstract}{Fortschr. Phys. {\bf 49}, 1117 (2001)}.
	\bibitem{Dodonov2000} V. V. Dodonov, O. V. Man'ko, V. I. Man'ko, and A. W\"{u}nsche, {\it Hilbert-Schmidt distance and non-classicality of states in quantum optics}, \href{http://dx.doi.org/10.1080/09500340008233385}{J. Mod. Opt. {\bf 47}, 633 (2000)}.
	\bibitem{Marian2002} P. Marian, T. A. Marian, and H. Scutaru, {\it Quantifying Nonclassicality of One-Mode Gaussian States of the Radiation Field}, \href{http://link.aps.org/doi/10.1103/PhysRevLett.88.153601}{Phys. Rev. Lett. {\bf 88}, 153601 (2002)}.
	\bibitem{Marian2004} P. Marian, T. A. Marian, and H. Scutaru, {\it Distinguishability and nonclassicality of one-mode Gaussian states}, \href{http://link.aps.org/doi/10.1103/PhysRevA.69.022104}{Phys. Rev. A {\bf 69}, 022104 (2004)}.
	\bibitem{Sperling2015} J. Sperling and W. Vogel, {\it Convex ordering and quantification of quantumness}, \href{http://stacks.iop.org/1402-4896/90/i=7/a=074024}{Phys. Scr. {\bf 90}, 074024 (2015)}.
	\bibitem{Lee1991} C. T. Lee, {\it Measure of the nonclassicality of nonclassical states}, \href{http://link.aps.org/doi/10.1103/PhysRevA.44.R2775}{Phys. Rev. A {\bf 44}, R2775(R) (1991)}.
	\bibitem{Miranowicz2015} I. I. Arkhipov, J. Pe\v{r}ina Jr., J. Pe\v{r}ina, and A. Miranowicz, {\it Comparative study of nonclassicality, entanglement, and dimensionality of multimode noisy twin beams}, \href{http://link.aps.org/doi/10.1103/PhysRevA.91.033837}{Phys. Rev. A {\bf 91}, 033837 (2015)}.
	\bibitem{Luetkenhaus1995} N. L\"utkenhaus and S. M. Barnett, {\it Nonclassical effects in phase space}, \href{http://link.aps.org/doi/10.1103/PhysRevA.51.3340}{Phys. Rev. A {\bf 51}, 3340 (1995)}.
	\bibitem{Kenfack2004} A. Kenfack and K. \.{Z}yczkowski, {\it Negativity of the Wigner function as an indicator of non-classicality}, \href{http://stacks.iop.org/1464-4266/6/i=10/a=003}{J. Opt. B {\bf 6}, 396 (2004)}.
%	
	\bibitem{Aharonov1966} Y. Aharonov, D. Falkoff, E. Lerner, and H. Pendleton, {\it A quantum characterization of classical radiation}, \href{http://dx.doi.org/10.1016/0003-4916(66)90079-0}{Ann. Phys. (N.Y.) {\bf 39}, 498 (1966)}.
	\bibitem{Kim2002} M. S. Kim, W. Son, V. Bu\v{z}ek, and P. L. Knight, {\it Entanglement by a beam splitter: Nonclassicality as a prerequisite for entanglement}, \href{http://link.aps.org/doi/10.1103/PhysRevA.65.032323}{Phys. Rev. A {\bf 65}, 032323 (2002)}.
	\bibitem{Wang2002} W. Xiang-bin, {\it Theorem for the beam-splitter entangler}, \href{http://link.aps.org/doi/10.1103/PhysRevA.66.024303}{Phys. Rev. A {\bf 66}, 024303 (2002)}.
	\bibitem{Miranowicz2016} I. I. Arkhipov, J. Pe\v{r}ina Jr., J. Svozil\'{\i}k, and A. Miranowicz, {\it Nonclassicality Invariant of General Two-Mode Gaussian States}, \href{http://dx.doi.org/10.1038/srep26523}{Sci. Rep. {\bf 6}, 26523 (2016)}.
	\bibitem{Arkhipov2016} I. I. Arkhipov, J. Pe\v{r}ina Jr., J. Pe\v{r}ina, and A. Miranowicz, {\it Interplay of nonclassicality and entanglement of two-mode Gaussian fields generated in optical parametric processes}, \href{http://link.aps.org/doi/10.1103/PhysRevA.94.013807}{Phys. Rev. A {\bf 94}, 013807 (2016)}.
%	
	\bibitem{Gehrke2012} C. Gehrke, J. Sperling, and W. Vogel, {\it Quantification of nonclassicality}, \href{http://link.aps.org/doi/10.1103/PhysRevA.86.052118}{Phys. Rev. A {\bf 86}, 052118 (2012)}.
	\bibitem{Vogel2014} W. Vogel and J. Sperling, {\it Unified quantification of nonclassicality and entanglement}, \href{http://link.aps.org/doi/10.1103/PhysRevA.89.052302}{Phys. Rev. A {\bf 89}, 052302 (2014)}. 
	\bibitem{Sanpera2001} A. Sanpera, D. Bru\ss{}, and M. Lewenstein, {\it Schmidt-number witnesses and bound entanglement}, \href{http://link.aps.org/doi/10.1103/PhysRevA.63.050301}{Phys. Rev. A {\bf 63}, 050301(R) (2001)}.
	\bibitem{Ternal2000} B. M. Terhal and P. Horodecki, {\it Schmidt number for density matrices}, \href{http://link.aps.org/doi/10.1103/PhysRevA.61.040301}{Phys. Rev. A {\bf 61}, 040301(R) (2000)}.
%	\bibitem{Sperling2015} J. Sperling and W. Vogel, {\it Convex ordering and quantification of quantumness}, \href{http://stacks.iop.org/1402-4896/90/i=7/a=074024}{Phys. Scr. {\bf 90}, 074024 (2015)}.
	\bibitem{Mraz2014} M. Mraz, J. Sperling, W. Vogel, and B. Hage, {\it Witnessing the degree of nonclassicality of light}, \href{http://link.aps.org/doi/10.1103/PhysRevA.90.033812}{Phys. Rev. A {\bf 90}, 033812 (2014)}.
	\bibitem{Bochner1933} S. Bochner, {\it Monotone Funktionen, Stieltjessche Integrale und harmonische Analyse}, \href{http://link.springer.com/article/10.1007/BF01452844}{Mathematische Annalen {\bf 108}, 378 (1933)}. %German grammar in the title
%	
	\bibitem{Vogel2000} W. Vogel, {\it Nonclassical states: An observable criterion}, \href{http://link.aps.org/doi/10.1103/PhysRevLett.84.1849}{Phys. Rev. Lett. {\bf 84}, 1849 (2000)}.
	\bibitem{Richter2002} Th. Richter and W. Vogel, {\it Nonclassicality of quantum states: A hierarchy of observable conditions}, \href{http://link.aps.org/doi/10.1103/PhysRevLett.89.283601}{Phys. Rev. Lett. {\bf 89}, 283601 (2002)}.

	\bibitem{Sperling2016} J. Sperling, {\it Characterizing maximally singular phase-space distributions}, \href{http://link.aps.org/doi/10.1103/PhysRevA.94.013814}{Phys. Rev. A {\bf 94}, 013814 (2016)}.
	\bibitem{Cahill1969} K. E. Cahill, {\it Pure States and the $P$ Representation}, \href{http://link.aps.org/doi/10.1103/PhysRev.180.1239}{Phys. Rev. {\bf 180}, 1239 (1969)}.
	\bibitem{Rahimi2013} S. Rahimi-Keshari, T. Kiesel, W. Vogel, S. Grandi, A. Zavatta, and M. Bellini, {\it Quantum Process Nonclassicality}, \href{http://link.aps.org/doi/10.1103/PhysRevLett.110.160401}{Phys. Rev. Lett. {\bf 110}, 160401 (2013)}.
% 	
	\bibitem{Bellini2007} V. Parigi, A. Zavatta, M. Kim, M. Bellini, {\it Probing Quantum Commutation Rules by Addition and Subtraction of Single Photons to/from a Light Field}, \href{http://science.sciencemag.org/content/317/5846/1890}{Science {\bf 317}, 1890 (2007)}.
	\bibitem{Zavatta2007} A. Zavatta, V. Parigi, and M. Bellini, {\it Experimental nonclassicality of single-photon-added thermal light states}, \href{http://link.aps.org/doi/10.1103/PhysRevA.75.052106}{Phys. Rev. A {\bf 75}, 052106 (2007)}.
	\bibitem{Kiesel2008} T. Kiesel, W. Vogel, V. Parigi, A. Zavatta, and M. Bellini, {\it Experimental determination of a nonclassical Glauber-Sudarshan $P$ function}, \href{http://link.aps.org/doi/10.1103/PhysRevA.78.021804}{Phys. Rev. A {\bf 78}, 021804(R) (2008)}.
	\bibitem{Fiurasek2009} J. Fiur\'a\v{s}ek, {\it Engineering quantum operations on traveling light beams by multiple photon addition and subtraction}, \href{http://link.aps.org/doi/10.1103/PhysRevA.80.053822}{Phys. Rev. A {\bf 80}, 053822 (2009)}.
	\bibitem{Dodonov2009} A. V. Dodonov and S. S. Mizrahi, {\it Smooth quantum-classical transition in photon subtraction and addition processes}, \href{http://link.aps.org/doi/10.1103/PhysRevA.79.023821}{Phys. Rev. A {\bf 79}, 023821 (2009)}.
%
 	\bibitem{Sperling2014} J. Sperling, W. Vogel, and G. S. Agarwal, {\it Quantum state engineering by click counting}, \href{http://link.aps.org/doi/10.1103/PhysRevA.89.043829}{Phys. Rev. A {\bf 89}, 043829 (2014)}.
	\bibitem{Lvovsky2002} A. I. Lvovsky and J. H. Shapiro, {\it Nonclassical character of statistical mixtures of the single-photon and vacuum optical states}, \href{http://link.aps.org/doi/10.1103/PhysRevA.65.033830}{Phys. Rev. A {\bf 65}, 033830 (2002)}.
	\bibitem{Kiesel2009} T. Kiesel, W. Vogel, B. Hage, J. DiGuglielmo, A. Samblowski, and R. Schnabel, {\it Experimental test of nonclassicality criteria for phase-diffused squeezed states}, \href{http://link.aps.org/doi/10.1103/PhysRevA.79.022122}{Phys. Rev. A {\bf 79}, 022122 (2009)}.
	\bibitem{Ryl2015} S. Ryl, J. Sperling, E. Agudelo, M. Mraz, S. K\"ohnke, B. Hage, and W. Vogel, {\it Unified nonclassicality criteria}, \href{http://link.aps.org/doi/10.1103/PhysRevA.92.011801}{Phys. Rev. A {\bf 92}, 011801(R) (2015)}.
	\bibitem{Leonhardt1995} U. Leonhardt, H. Paul, and G. M. D’Ariano, {\it Tomographic reconstruction of the density matrix via pattern functions}, \href{http://link.aps.org/doi/10.1103/PhysRevA.52.4899}{Phys. Rev. A 52, 4899 (1995)}.
	\bibitem{Hillery1985} M. Hillery, {\it Classical pure states are coherent states}, \href{http://www.sciencedirect.com/science/article/pii/0375960185904839}{Phys. Lett. A {\bf 111}, 409 (1985)}.
%
	\bibitem{KieselThesis} T. Kiesel, {\it Verification of nonclassicality in phase space}, \href{http://rosdok.uni-rostock.de/resolve?urn=urn:nbn:de:gbv:28-diss2011-0145-7&pdf}{PhD thesis, Universit\"at Rostock, 2011}.
	\bibitem{Walmsley2008} P. J. Mosley, J. S. Lundeen, B. J. Smith, P. Wasylczyk, A. B. U'Ren, C. Silberhorn, and I. A. Walmsley, {\it Heralded generation of ultrafast single photons in pure quantum states}, \href{http://link.aps.org/doi/10.1103/PhysRevLett.100.133601}{Phys. Rev. Lett. {\bf 100}, 133601 (2008)}.
	\bibitem{Smith2013} M. Cooper, L. J. Wright, C. S\"{o}ller, and B. J. Smith, {\it Experimental generation of multi-photon Fock states}, \href{http://www.opticsexpress.org/abstract.cfm?URI=oe-21-5-5309}{Opt. Express {\bf 21}, 5309 (2013)}.
%	\bibitem{Krumm2016} F. Krumm, J. Sperling, and W. Vogel, {\it Multitime correlation functions in nonclassical stochastic processes}, \href{http://link.aps.org/doi/10.1103/PhysRevA.93.063843}{Phys. Rev. A {\bf 93}, 063843 (2016)}.
% 	\bibitem{Dodonov1995} V. V. Dodonov, V. I. Man'ko, and D. E. Nikonov, {\it Even and odd coherent states for multimode parametric systems}, Phys. Rev. A {\bf 51}, 3328 (1995).
% 	\bibitem{Korennoy2011} Ya. A. Korennoy and V. I. Man'ko, {\it Optical tomography of photon-added coherent states, even and odd coherent states, and thermal states}, Phys. Rev. A {\bf 83}, 053817 (2011).
\end{thebibliography}
\end{document}